# Deep Reinforcement Learning for Backhaul Link Selection for Network Slices in IAB Networks

António J. Morgado†, Firooz B. Saghezchi*, Pablo Fondo-Ferreiro‡, Felipe Gil-Castiñeira‡, Maria Papaioannou⟂, Kostas Ramantas§, Jonathan Rodriguez†

†*Faculty of Computing, Engineering and Science, University of South Wales, Pontypridd CF37 1DL, U.K.*
\**Instituto de Telecomunicações, Universidade de Aveiro, Aveiro, Portugal*
‡*Information Technologies Group, atlanTTic Research Center, University of Vigo, Vigo, Spain*
⟂*Faculty of Engineering and Science, University of Greenwich, Chatham Maritime, U.K.*
§*Iquadrat R&D Department, Iquadrat Informatica, S.L, Barcelona, Spain*
{antonio.dasilvamorgado, jonathan.rodriguez}@southwales.ac.uk, firooz@av.it.pt, {pfondo,xil}@gti.uvigo.es, m.papaioannou@greenwich.ac.uk, kramantas@iquadrat.com

*Abstract*—Integrated Access and Backhaul (IAB) has been recently proposed by 3GPP to enable network operators to deploy fifth generation (5G) mobile networks with reduced costs. In this paper, we propose to use IAB to build a dynamic wireless backhaul network capable to provide additional capacity to those Base Stations (BS) experiencing congestion momentarily. As the mobile traffic demand varies across time and space, and the number of slice combinations deployed in a BS can be prohibitively high, we propose to use Deep Reinforcement Learning (DRL) to select, from a set of candidate BSs, the one that can provide backhaul capacity for each of the slices deployed in a congested BS. Our results show that a Double Deep Q-Network (DDQN) agent using a fully connected neural network and the Rectified Linear Unit (ReLU) activation function with only one hidden layer is capable to perform the BS selection task successfully, without any failure during the test phase, after being trained for around 20 episodes.

*Keywords—machine learning, deep reinforcement learning, integrated access and backhaul, resource allocation, backhaul link selection, network slicing.*

## I. INTRODUCTION

Integrated Access and Backhaul (IAB) is a feature of fifth Generation (5G) mobile networks that enables Base Stations (BS) with wired backhaul, called IAB donors, to provide wireless backhaul links to the other ones either with congested or with no wired backhaul links at all, called IAB nodes, using the free channels in the access band (in-band) or using a different band (out-of-band). The IAB feature was first introduced by 3GPP in Release 16. It was enhanced in Release 17 with inter-donor migration and topological redundancy, and in Release 18 with the introduction of mobile IAB-nodes mounted on vehicles.

The main use cases envisioned for IAB are coverage extension, deployment of outdoor small cells, providing Fixed Wireless Access (FWA) for indoor hotspots [1] or alternative backhaul links for a BS with congested wired backhaul. However, in these use cases, one of the major challenges is to select, in each time instant, among the neighboring BSs, the best one that can provide the additional capacity, according to the variations in the traffic demand of all IAB BSs (including the donors and the recipient), and without interfering with any incumbent network that operates on the same band in the vicinities. In this sense, IAB can also be considered as a spectrum sharing scheme since the spectrum band used for access purposes can be used both to provide connectivity to UEs and to provide wireless backhaul to nearby BSs, as long as any of these connections does not cause harmful interference to any incumbent network operating in the same band and in the same region.

In addition to IAB, network slicing is another main 5G feature, which enables to create multiple isolated logical network slices over the same physical infrastructure. This allows different tenants to share the infrastructure and offer different and independent services to their customers, which can lead to a considerably reduced cost for the operators and the end users. Furthermore, network slicing allows software-based network reconfiguration for an end-to-end and cross domain network and service management, which can better cope with end users' Quality-of-Service (QoS) requirements.

In this paper, we investigate in-band IAB to dynamically provide additional backhaul capacity to the network slices of a congested BS through directional beams deployed in the same band as the access network. Since the mobile traffic demand may vary tremendously both in time and space [2], we employ Deep Reinforcement Learning (DRL) to dynamically select, at each time instant, the BS(s) that can provide wireless backhaul link(s) to the network slice(s) of the congested BS without impacting the surrounding BSs and their served slices. That is, we use network slicing for backhaul purposes and assume that there can be as many backhaul beams as network slices, each one pointed towards a candidate BS selected for a specific slice(s). To the best of our knowledge, this is the first work applying DRL for resource management at network slice level in IAB networks. Previous contributions incorporated DRL for either IAB resource management or network slicing management, but not both. As such, the main contributions of this work can be summarized as follows:

- We propose to use in-band IAB for backhauling the network slices of a congested BS, without disturbing the network slices served by the donor IAB BS.
- We construct and validate a DRL model to select the best IAB BS from the set of candidate BSs for each network slice.

This work was supported in part by Xunta de Galicia (Spain) under grant ED481B-2022-019 and by EXPLOR project funded by H2020-MSCA-RISE-2019 (grant agreement ID: 872897).

The rest of this paper is organized as follows: in Section II, we provide an overview of the state of the art on applying DRL either for IAB resource management or for slicing management. In Section III, we describe our system model, while in Section IV, we present our DRL model for IAB-donor selection. In section V, we present our simulation environment and the validation results. Finally, in Section VI, we conclude the paper and draw guidelines for future work.

## II. RELATED WORK

IAB is considered as a means to reduce deployment costs in 5G networks and beyond, especially in ultra-dense scenarios such as millimeter wave (mmWave) networks [3]. In IAB, wireless backhaul links between BSs are used to transport access traffic. The main challenges in IAB relate to the self-configuration of the network and the traffic path selection in order to guarantee the desired QoS and optimize network resource utilization.

Most of the existing works on IAB focus on RAN resource allocation. Due to the complexity of the scenarios, DRL techniques have been explored for addressing these challenges. A DRL-based radio resource management solution for congestion avoidance has been proposed in [4]. The authors in [5] used DRL-based techniques for spectrum allocation with the aim of maximizing the sum log-rate of the users. DRL has also been used in [6] for jointly addressing the spectrum allocation and power control in IAB networks. The authors in [7] proposed a DRL-based cross-layer approach for jointly tackling routing and radio resource allocation in multi-hop IAB networks.

The authors in [8] propose an autonomous and distributed approach for user association in multi-hop IAB networks based on Reinforcement Learning (RL), bandwidth partitioning and reactive load balancing depending on the load of the neighboring BSs. Spectrum allocation in mmWave IAB has also been addressed in [9], where a prediction of user behavior is used to jointly optimize energy consumption and spectrum efficiency, using Deep Recurrent Q-Network (DRQN), which integrates a Long Short-Term Memory (LSTM) recurrent neural network with DRL. In [10], the authors propose a DRL-based scheme for adapting backhaul according to the load on the access network, enhanced with a Recursive Discrete Choice Model (RDCM). Their proposed scheme outperforms baseline strategies (i.e., conventional DRL without RDCM, and Generative Model-Based Learning (GMBL)) in terms of throughput and delay. DRL has also been recently proposed in IAB for topology design and planning [11], providing a sub-optimal but scalable solution.

DRL has also been explored for addressing resource management at network slice level [12], both for Radio Access Network (RAN) and core network slicing. The authors in [13] use DRL for network slice reconfiguration in order to meet the required Service Level Agreements (SLAs) of the slices and maintain high resource efficiency. Due to the large dimensionality of the action space, they propose the use of a branch dueling Q-network algorithm. An autonomous virtual resource slicing framework is proposed in [14], where DRL is employed to adapt the amount of resources assigned to each network slice, showing an overall improvement in the resource usage and QoS satisfaction.

However, despite these previous efforts, the application of DRL in IAB networks is mostly limited to addressing RAN resource allocation problem. In this paper, we extend the existing works by focusing on resource management for IAB at the network slice level. In particular, as mentioned above, we incorporate DRL for selecting IAB donor BS(s) for backhauling network slices of a congested BS.

## III. SYSTEM MODEL

In this section, we present our addressed IAB scenario for selecting the donor IAB BS for each network slice of a congested BS, including network topology, the assumed network slices for the congested BS and the traffic profiles for the surrounding (donor) IAB BSs.

### A. Addressed Scenario

Fig. 1 illustrates our addressed scenario, where all the seven 5G BSs (gNodeBs) have a wired (fiber optic) backhaul connecting them to the core network, and may also use IAB to borrow additional backhaul capacity from the neighboring BSs. We assume that the congested BS, i.e., the BS needing to borrow capacity from the neighboring BSs is BS1. Different aspects that need to be considered in this scenario include: the network topology, the QoS requirements of the $S$ slices to be served by BS1, and the traffic load in the surrounding BSs.

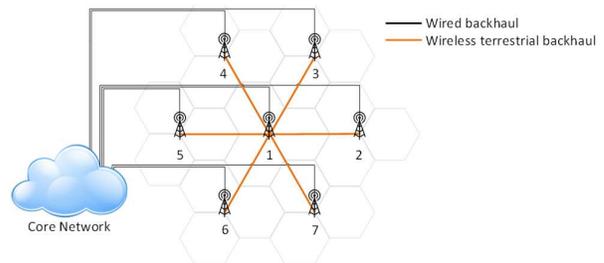

Fig. 1. IAB scenario under study.

### 1) Network Topology

As illustrated by Fig. 1, considering $N$=7 BSs, our backhaul network is composed of $N+1$ nodes, where $N$ nodes (nodes $1…N$) represent the IAB BSs themselves and the remaining node (node 0) represents the core network. Each node is characterized by a name and its geographical location. The nodes are connected together by directional links represented by the tuples:

$$Link(x, i, j, b) \quad (1)$$

where $x \in \{wired, wireless\}$ represents the link type, $i \in \{0,1,\cdots,N\}$ represents the source node and $j \in \{0,1,\cdots,N\}$ represents the target node and $b$ represents the total bandwidth of the link. In the case of wired links, we assume there are 2 fibers, each one to transmit traffic in one direction (DL, UL). On the other hand, in the case of wireless links, we suppose that the links operate in orthogonal resources (in space domain) and hence do not interfere with each other or with any incumbent network that might be operating in the same band in the same region.

### 2) Traffic Profiles of Network Slices Served by BS1

The traffic profiles of the $S$ network slices served by BS1 define the time variation of the traffic demand, required in each direction (DL, UL), by each of these slices during one day, discretized in 15-minute intervals. The $S$ traffic profiles are stored in a timetable with the following format:

$$Slice\_profile(t, i, s, thdl, thul) \quad (2)$$

where *t* is the time, which spans a 24h period, discretized in 15-minute intervals, *i* is the node transmitting that slice, *s* is the slice identifier. For each time interval, *thdl* and *thul* represent the throughput required in that time interval by slice *s* in DL and UL, respectively.

*3) Traffic Loads of the Remaining N-1 BSs*

We define several traffic load profiles to represent BSs in different situations, e.g., a BS periodically congested, a BS congested in one part of the day, and a BS not congested. Each profile is stored in a different timetable with the following format:

$$BS\_load\_profile(t, thdl, thul) \quad (3)$$

where *t* is the time, which spans a period of 24h discretized in 15-minute intervals. For each 15-min interval, the load profile defines the total throughput supported by the BS in DL and UL, respectively. Each of these defined profiles can then be assigned to any of the *N-1* BSs following any assignment strategy. We will discuss this further in Section V-D.

## IV. PROPOSED DRL MODEL FOR IAB BS SELECTION

The task of the DRL agent is to decide, every 15 minutes, if the BS1 needs to borrow wireless backhaul capacity from the neighboring *N-1* BSs for any of its *S* served slices. To do so, the agent will observe the state of the (network) environment, select an action, and apply it to collect a reward that measures how good the selected action was.

*A. Action Model*

In Fig. 1, the action that the DRL agent has to perform is to select, every 15 minutes, the *S* backhaul links (UL, DL) for each of the *S* network slices, taking into account the current traffic load of the wired connection of the BS1 and the traffic load of the wireless connections of the surrounding BSs (BS2…BS7).

Table I shows the action vector, where we have seven possible options. We use this format for the action because the neural network that approximates the critic function Q(s,a) can have a different input value for each backhaul option (BS1,…BS7), thus more effectively distinguishing the different actions.

TABLE I. FORMAT OF THE ACTION VECTOR

| Selectable links | | | | | | |
|---|---|---|---|---|---|---|
| Wired | Wireless | | | | | |
| BS1 | BS2 | BS3 | BS4 | BS5 | BS6 | BS7 |
| Logical | Logical | Logical | Logical | Logical | Logical | Logical |
| Only one of these links can be TRUE at a time | | | | | | |

In this work we assume that UL and DL traffic of a given slice will be carried by the same backhaul link. Moreover, the DRL agent makes one decision at a time, i.e., to allocate the links for the *S* slices, it need to perform a sequence of *S* actions. Thus, the action space is of discrete type, composed of length-N binary vectors whose elements are all zero except one (i.e., the selected IAB BS); therefore, the size of the action space is N.

*B. Observation Model*

In the observation, we include the throughput requirements of the slice being backhauled in a given time, i.e., the QoS level required by that network slice. Besides this information, in a second part, the observation also includes the information about the bandwidth currently available in the wired backhaul link of BS1 and in the wireless backhaul links offered by each of the surrounding IAB BSs (BS2…BS7). All this information is included in an observation vector as illustrated in Table II.

TABLE II. FORMAT OF THE OBSERVATION VECTOR

| Slice s requirements. | | Free bandwidth (Mbps) | | | | | |
|---|---|---|---|---|---|---|---|
| Throughput | | BS1 | BS2 | … | BS7 | | |
| DL | UL | DL | UL | DL | UL | … | DL | UL |

All these observation values are normalized so they all vary in the same interval. The objective of doing this was to ensure that all inputs have the same impact when they are provided to the neural network.

*C. Reward Model*

The reward model defines the objective of the DRL agent. We adopted a reward model where the agent receives a reward +1 when it selects a link that connects BS1 to the core network with the required DL and UL QoS so it can be used as the backhaul for the slice under consideration. Otherwise, i.e. if the selected link cannot provide the required QoS level to the network slice, the agent receives a reward 0, and no link is allocated for that slice. This reward model is effective and at the same time allows us to easily calculate the maximum reward that the DRL agent can collect over one episode. This is important to judge the performance of the DRL agent during training, cross validation, and testing.

## V. PERFORMANCE EVALUATION RESULTS

*A. Simulator Setup*

We implemented a custom made simulator in a way that the DRL agent interacts with a Software-Defined Network (SDN) simulator, which represents the environment, as depicted in Fig. 2. This environment sends commands to the SDN simulator whenever it wants to allocate a backhaul link between BS1 and the core network. Then, the environment reads from the SDN simulator what was the bandwidth effectively allocated for each slice, computes the reward and the next state, and returns them to the DRL agent.

In fact, the SDN backhaul simulator acts as part of the environment for the DRL agent. It enables the execution of the actions selected by the DRL agent on the backhaul network and observing the resulting state changes. The actions correspond to the selection of a specific backhaul link to transport the network traffic for a particular slice, provided that there is enough capacity available in the selected path up to the core network. The observations provide information about network KPIs, e.g., the actual throughput perceived by the network slices.

The SDN backhaul architecture consists of a set of SDN switches co-located along with the IAB BSs, interconnected with the different connectivity options. The SDN switches are connected to the SDN controller. An SDN application, running on top of the SDN controller, is responsible for reconfiguring the flow tables of the SDN switches according to the desired forwarding path for each network slice.

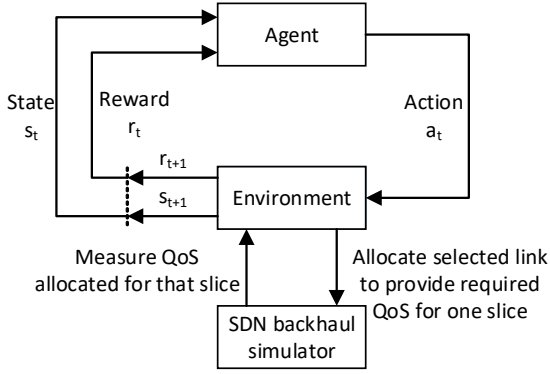

Fig. 2. Simulator architecture

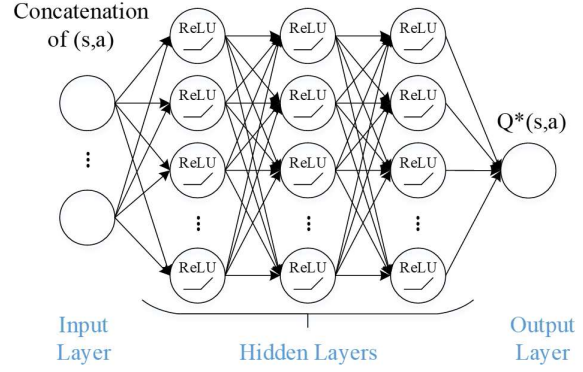

Fig. 3. Fully-connected artificial neural network with ReLU activation for modelling the critic function

The network topology (Fig. 1) is internally represented as a directed graph, with a special node labelled as the core network (node 0). Each edge of the graph represents a backhaul link and contains the available capacity (throughput) over that link. When the DRL agent selects a backhaul link for a given slice, the observed throughput measured for that slice will be equal to the requested throughput if that link is allocated, i.e., if there is enough capacity available in that link to accommodate the slice. Otherwise, if there is not enough capacity to accommodate the slice in the selected link, the slice will not be allocated to any backhaul link and the measured allocated throughput for the slice will be 0.

*B. Reinforcement Learning Agent*

We use a Double Deep Q-Network (DDQN) [15] agent to select the backhaul links, since it works with continuous observation space and discrete action space. This model-free, value-based DRL agent works by estimation of the optimum state-action value function $Q^*(s,a)$, as given by (4), using two identical neural networks. As for the neural network, we use a fully connected network as depicted in Fig. 3. The optimum policy is then derived by the agent by selecting action $a$ that maximizes $Q^*(s,a)$ for a given state $s$.

$$Q^*(s,a) = \mathbb{E}\left\{r_{t+1} + \gamma \max_{a'} Q^*(s_{t+1}, a')\right\} \quad (4)$$

In Fig. 3, the number of input values is equal to the sum of the lengths of the observation vector (16 parameters, cf. Table II) and the action vector (7 parameters, cf. Table I), which adds up to 23 neurons. We concatenate the observation and action vectors and provide the concatenated vector as input to the neural network.

Each hidden layer is a fully connected layer with Rectified Linear Unit (ReLU) activation function. The output layer is similarly a fully connected layer, but with only one neuron – delivering the $Q^*(s,a)$ estimate, as shown in Fig. 3.

*C. Traffic Profiles of the Network Slices in BS 1*

As stated before, in our addressed scenario (Fig. 1), all the *seven 5G BSs* have a wired backhaul connecting them to the core network, and they also use IAB to borrow additional backhaul capacity from the neighboring BSs when needed.

The wired backhaul links have a bandwidth of 1Gbps (DL) / 1Gbps (UL), while the IAB links have a total bandwidth of 1Gbps (DL) / 1Gbps (UL) that is shared between the access and backhaul links. In other words, when an IAB BS has no local (access network) traffic, the entire 1Gbps (DL) / 1Gbps (UL) bandwidth will be available for wireless backhaul links. Otherwise, for instance, when the BS is heavy loaded, part of this bandwidth will be occupied by access network traffic and the remaining part by backhaul traffic.

We assume that the congested BS, i.e., the one needing to borrow wireless backhaul capacity from the neighboring BSs is BS1. For the purpose of simulation, we assume that during a typical day, this BS has to support three network slices with the throughput requirements depicted in Fig. 4. As we can see from the figure, from 05:30 to 12:30, the 1Gbps wired connection lacks enough capacity to support all the UL traffic. The same problem exists in DL from 14:30 to 18:30. Therefore, in these periods, the DRL agent has to select one of the neighboring BSs to wirelessly backhaul one or more of its served slices.

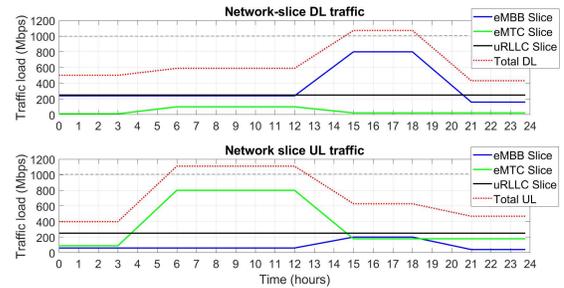

Fig. 4. Daily traffic of the network slices in BS1

*D. Traffic Profiles of the Surrounding BS*

The surrounding IAB BSs (BS2 to BS7) have to primarily carry the access network traffic for their associated UEs, and for this reason, they can only lend to BS1 the remaining capacity for wireless backhauling. Fig. 5 indicates the assumed load of the surrounding BSs in our simulations, where we assign profile *p* to BS number *bs* according to:

$$p=((bs-1) \bmod 3) + 1, \quad b=2,\ldots,7 \quad (5)$$

As a result, BS4 and BS7 end up having *profile 1*, which is heavily loaded only for small periods of time, BS2 and BS5 end up having *profile 2*, which is partially loaded, and BS3 and BS6 are heavy loaded mainly at noon.

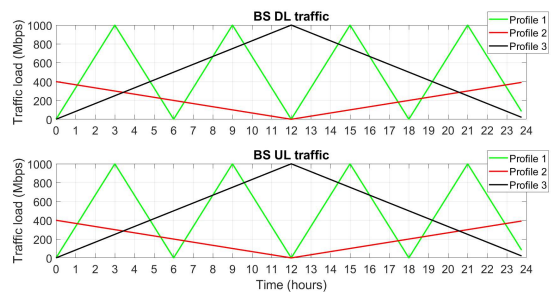

Fig. 5. Daily traffic load of surrounding BSs (BS2-7)

Therefore, the task of the DRL agent is to decide, every 15 minutes, if BS1 needs to borrow wireless backhaul capacity from the neighboring BSs for any of its 3 slices. In the affirmative case, i.e. during the periods 05:30-12:30 and 14:30-18:30, the DRL agent has to decide:

- which slice is going to have its backhaul traffic transmitted wirelessly;
- which BS is going to transmit this backhaul traffic for BS1.

To take these decisions, the agent will observe the state of the environment, select an action, apply it to the environment, and collect a reward that acts as feedback to evaluate how good the selected action was.

### E. Simulator validation

Since our reward model assigns a reward +1 every 15 minutes when the SDN simulator assigns a backhaul link for each slice, and a reward 0 otherwise, then the maximum undiscounted episode reward in each episode (24 hours) is given by:

$$3 \text{ (slices)} \times 24 \text{ (hours)} \times 60 \text{ (min)} / 15 \text{ (min)} = 288. \quad (6)$$

We perform simulations with different hyperparameters to fine tune the DDQN agent and have its episode reward converged to 288, as illustrated by Fig. 6. The best hyperparameters are described in Table III.

TABLE III. DRL HYPERPARAMETERS

| Hyperparameter | Value |
|---|---|
| $\alpha$ (learning rate) | 0.0001 |
| $\gamma$ (discount factor) | 0.99 |
| $\varepsilon$ (initial exploration probability) | 0.99 |
| $\varepsilon_{decay}$ ($\varepsilon$ decay rate) | 0.01 |
| M (minibatch size) | 64 samples |
| N (size of experience replay buffer) | 10000 samples |
| C (periodic update of the target critic) | 4 timesteps |
| Gradient descent optimization algorithm | adam (adaptive moment estimation) |

### F. Simulation results

In our simulations, each episode contains 288 samples as given by (6), of which, we use 70% (201 samples) for training, 10% (27 samples) for validation, and 20% (60 samples) for testing. In the training phase, we opt for early stopping to avoid overfitting. Hence, we stop the training when the moving average of the episode reward, for a window size of 10, reaches 97.5% of the maximum achievable value over a training episode, i.e., 0.975 * 201 = 195.975. We repeat the simulations using the critic network with 1, 3 and 5 fully connected hidden layers and ReLU activations and with 8, 16, 24, 32, 40 and 48 neurons per hidden layer. For each of these combinations, the number of episodes necessary to achieve the target episode reward (i.e., 97.5% of the maximum attainable value) is given by Table IV.

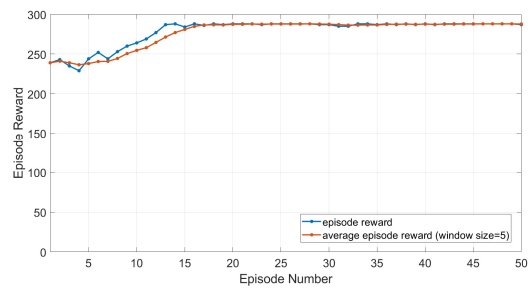

Fig. 6. Simulator achieves the optimum episode reward (288) steadily

TABLE IV. NUMBER OF TRAINING EPISODES TO REACH THE TARGET AVERAGE EPISODE REWARD

| No. of hidden layers | Neurons per layer | Training episodes | Last episode reward | Average episode reward | Validation reward | Testing reward |
|---|---|---|---|---|---|---|
| 5 | 48 | 12 | 201 | 196.7 | 27 | 60 |
| 5 | 40 | 13 | 201 | 196.0 | 27 | 60 |
| 5 | 32 | 12 | 198 | 200.1 | 27 | 60 |
| 5 | 24 | 17 | 199 | 197.7 | 26 | 59 |
| 5 | 16 | 23 | 201 | 197.2 | 27 | 60 |
| 5 | 8 | 86 | 196 | 198.1 | 27 | 60 |
| 3 | 48 | 11 | 201 | 198.4 | 26 | 59 |
| 3 | 40 | 11 | 201 | 198.4 | 27 | 60 |
| 3 | 32 | 17 | 199 | 197.7 | 27 | 60 |
| 3 | 24 | 34 | 201 | 196.3 | 18 | 57 |
| 3 | 16 | 34 | 201 | 196.6 | 27 | 60 |
| 3 | 8 | DNC | DNC | DNC | DNC | DNC |
| 1 | 48 | 30 | 198 | 196.0 | 26 | 59 |
| 1 | 40 | 48 | 201 | 196.0 | 27 | 60 |
| **1** | **32** | **21** | **201** | **196.8** | **27** | **60** |
| 1 | 24 | 52 | 201 | 196.7 | 27 | 60 |
| 1 | 16 | 17 | 201 | 196.9 | 27 | 58 |
| 1 | 8 | DNC | DNC | DNC | DNC | DNC |

DNC: did not converge

We observe from Table IV that the DDQN agent achieves a very good performance with a quite simple critic model. With only 1 hidden layer and 32 neurons per layer, it converges very fast (in 21 episodes) and never fails during the test phase (i.e., it collects the maximum attainable reward during the test phase, which is 60). This is important for not only the practical implementation feasibility, but the adaptation (retraining) of the agent to cope with potential variations in the environment (e.g., due to the varying traffic profiles of BS2-7). The episode reward collected during the training of this agent is shown in Fig. 7. We can verify that the simulation stopped because the average episode reward hit 97.5% of the maximum achievable episode reward after 21 episodes.

Fig. 8 shows the evolution of this DDQN agent's training (i.e., with one hidden layer and 32 neurons per hidden layer) in terms of the total throughput of BS1 (aggregated 3 slices, DL+UL). We see that the 3 slices requested an average traffic demand of 1,479 Gbps (the dashed constant line in the figure), which was first reached at around episode 15 and stabilized afterwards. Then, the training is stopped few episodes after, at episode 21.

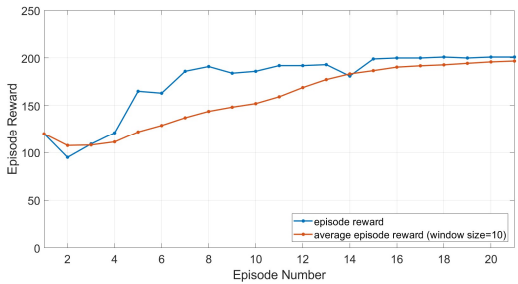

Fig. 7. Training of DDQN agent using a critic with 1 layer and 32 neurons

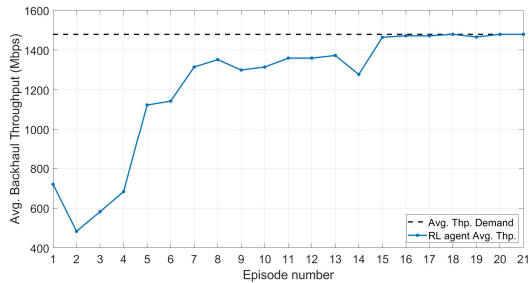

Fig. 8. Evolution of average throughput achieved during training

The fast training of the proposed DDQN model makes it interesting for practical implementation in real wireless environments where the channel conditions and traffic profiles may dynamically change. However, for practical implementation we need to have a testbed environment with IAB nodes with data collection functions to monitor the different links of the network and the delivered QoS for the slices, and to provide feedback to the RL agent. Moreover, to extend the proposed approach to larger topologies, we need to consider a conflict resolution mechanism for multi-agent RL system where each agent may compete with its neighbor agents for the wireless backhaul links.

## VI. CONCLUSIONS

In this paper, we exploited IAB to provide additional backhaul capacity to a congested BS to cope with the dynamic traffic load variations across time and space. We adopted a DRL approach and constructed a DDQN agent to select, among the candidate neighboring BSs, those that are capable to backhaul the traffic for each of the slices deployed in the congested BS. We conclude that a DDQN agent with a fully connected neural network and ReLU activation function, with a single hidden layer and 32 neurons per hidden layer is capable of performing the BS selection task with very high accuracy, after being trained for only 21 episodes. This shows that the agent can be implemented with minimum computation impact on the IAB BSs, and is able to cope with their dynamic traffic variations.

In a future work, we plan to study network slices whose QoS requirements are defined in terms of not only throughput, but also latency and assess the performance and computation complexity of the constructed DRL agent. In addition is also worth investigating the practical implementation of the proposed scheme in real testbeds and extend the work to larger topologies, e.g., using a multi-agent DRL approach.

REFERENCES

[1] O. Teyeb, A. Muhammad, G. Mildh, E. Dahlman, F. Barac and B. Makki, "Integrated Access Backhauled Networks," 2019 IEEE 90th Vehicular Technology Conference (VTC2019-Fall), Honolulu, HI, USA, 2019, pp. 1-5.

[2] A. J. Morgado, F. B. Saghezchi, S. Mumtaz, V. Frascolla, J. Rodriguez and I. Otung, "A Novel Machine Learning-Based Scheme for Spectrum Sharing in Virtualized 5G Networks," in IEEE Transactions on Intelligent Transportation Systems, vol. 23, no. 10, pp. 19691-19703, Oct. 2022,

[3] M. Polese, M. Giordani, T. Zugno, A. Roy, S. Goyal, D. Castor, and M. Zorzi, "Integrated access and backhaul in 5G mmWave networks: Potential and challenges,", in IEEE Communications Magazine, vol. 58, no. 3, pp. 62-68, March 2020.

[4] M. M. Sande, M. C. Hlophe, and B. T. Maharaj, "Access and radio resource management for IAB networks using deep reinforcement learning," in IEEE Access, vol. 9, pp. 114218-114234, August 2021.

[5] W. Lei, Y. Ye, and M. Xiao, "Deep reinforcement learning-based spectrum allocation in integrated access and backhaul networks," in IEEE Transactions on Cognitive Communications and Networking, vol. 6, no. 3, pp. 970-979, September 2020.

[6] Q. Cheng, Z. Wei, and J. Yuan, "Deep reinforcement learning-based spectrum allocation and power management for IAB networks," in 2021 IEEE International Conference on Communications Workshops (ICC Workshops), pp. 1-6, June 2021.

[7] H. Yin, S. Roy, and L. Cao, "Routing and Resource Allocation for IAB Multi-Hop Network in 5G Advanced," in IEEE Transactions on Communications, vol. 70, no. 10, pp. 6704-6717, October 2022.

[8] M. M. Sande, M. C. Hlophe and B. T. Maharaj, "Instantaneous Load-Based User Association in Multi-Hop IAB Networks using Reinforcement Learning," GLOBECOM 2020 - 2020 IEEE Global Communications Conference, Taipei, Taiwan, 2020, pp. 1-6, doi: 10.1109/GLOBECOM42002.2020.9322230.

[9] X. Zhou and X. Dong, "Intelligent Spectrum Allocation for mmWave Integrated Backhaul and Access Network", 2021 9th International Conference on Intelligent Computing and Wireless Optical Communications (ICWOC), pp. 69-74, 2021.

[10] M. M. Sande, M. C. Hlophe and B. T. S. Maharaj, "A Backhaul Adaptation Scheme for IAB Networks Using Deep Reinforcement Learning With Recursive Discrete Choice Model," in IEEE Access, vol. 11, pp. 14181-14201, 2023, doi: 10.1109/ACCESS.2023.3243519.

[11] A. Abdelmoaty, D. Naboulsi, G. Dahman, G. Poitau and F. Gagnon, "Resilient Topology Design for Wireless Backhaul: A Deep Reinforcement Learning Approach," in IEEE Wireless Communications Letters, vol. 11, no. 12, pp. 2532-2536, Dec. 2022, doi: 10.1109/LWC.2022.3207358.

[12] R. Li, Z. Zhao, Q. Sun, I. Chih-Lin, C. Yang, X. Chen, et al., "Deep reinforcement learning for resource management in network slicing", IEEE Access, vol. 6, pp. 74429-74441, 2018.

[13] F. Wei, G. Feng, Y. Sun, Y. Wang, S. Qin and Y. -C. Liang, "Network Slice Reconfiguration by Exploiting Deep Reinforcement Learning With Large Action Space," in IEEE Transactions on Network and Service Management, vol. 17, no. 4, pp. 2197-2211, Dec. 2020

[14] G. Sun, Z. T. Gebrekidan, G. O. Boateng, D. Ayepah-Mensah and W. Jiang, "Dynamic reservation and deep reinforcement learning based autonomous resource slicing for virtualized radio access networks", IEEE Access, vol. 7, pp. 45758-45772, 2019.

[15] H. van Hasselt, A. Guez, and D. Silver, "Deep reinforcement learning with double Q-learning," in Proc. of the Thirtieth AAAI Conference on Artificial Intelligence, vol. 30, no. 1, pp. 2094–2100, March 2016.